\documentclass[twocolumn,prb,reprint,superscriptaddress,floatfix,amsmath,amssymb,aps]{revtex4-2}
\pdfoutput=1

\usepackage[utf8]{inputenc}
\usepackage[caption=false,subrefformat=parens,labelformat=parens]{subfig}
\usepackage{graphicx}
\usepackage{float}
\usepackage{amsmath}
\usepackage{dcolumn}
\usepackage{bm}
\usepackage{xcolor}
\usepackage{xfrac}
\usepackage[colorlinks=true, allcolors=blue]{hyperref}
\usepackage{chemformula}
\usepackage{siunitx}

\usepackage[T1]{fontenc}
\usepackage{textcomp}

\newcommand{\mct}{(Mn$_{1-x}$Cr$_x$)Te}

\begin{document}
\title{Intertwined ferromagnetic and altermagnetic orders in Cr-doped MnTe}

\author{Jonathan M. DeStefano}
\altaffiliation{These authors contributed equally to this work.}
\affiliation{Department of Physics, University of Washington, Seattle, WA 98112, USA}
\author{Yaofeng Xie}
\altaffiliation{These authors contributed equally to this work.}
\affiliation{Department of Physics, University of Washington, Seattle, WA 98112, USA}
\author{Guodong Ren}
\affiliation{Materials Science and Engineering Department, University of Washington, Seattle WA, 98115, USA}
\author{Mai Nguyen}
\affiliation{Department of Physics, University of Washington, Seattle, WA 98112, USA}
\author{Yue Sun}
\affiliation{Department of Physics, University of Washington, Seattle, WA 98112, USA}
\author{Xiaodong Xu}
\affiliation{Department of Physics, University of Washington, Seattle, WA 98112, USA}
\affiliation{Materials Science and Engineering Department, University of Washington, Seattle WA, 98115, USA}
\author{Juan Carlos Idrobo}
\affiliation{Materials Science and Engineering Department, University of Washington, Seattle WA, 98115, USA}
\affiliation{Physical and Computational Sciences Directorate, Pacific Northwest National Laboratory, Richland WA, 99354, USA}
\author{Jiun-Haw Chu}
\email{jhchu@uw.edu}
\affiliation{Department of Physics, University of Washington, Seattle, WA 98112, USA}

\date{\today}

\begin{abstract}
Ferromagnetism and altermagnetism both break time-reversal symmetry, yet their distinct transformation properties under crystalline symmetries forbid direct bilinear coupling between their order parameters. Whether these two magnetic orders can remain thermodynamically distinct while strongly influencing one another remains an open question. Here we demonstrate the homogeneous coexistence of ferromagnetism and altermagnetism in Cr-doped MnTe. By tuning the ferromagnetic ordering temperature across the nearly unchanged Néel temperature with Cr doping, we uncover a striking interplay between the two orders: the ferromagnetic transition becomes a crossover when altermagnetism develops first, whereas the onset of altermagnetism strongly suppresses the out-of-plane ferromagnetic remanence when ferromagnetism develops first. Moreover, spin–orbit coupling enables symmetry-selective interactions through which the ferromagnetic moment controls the altermagnetic domain manifold, with its out-of-plane and in-plane components biasing the time-reversal ($Z_2$) and orientational ($Z_3$) sectors, respectively. Our results reveal how two symmetry-distinct magnetic orders can retain separate ordering tendencies while remaining strongly coupled, providing a route towards internal control of altermagnetic domain states.

\end{abstract}

\maketitle

Altermagnetism has recently emerged as a distinct form of magnetic order~\cite{ma2021multifunctional,smejkal2022AMreview1,smejkal2022AMreview2,song2025AMreview,jungwirth2026symmetry}. Like conventional antiferromagnets, altermagnets possess compensated magnetic moments and vanishing net magnetization. Unlike conventional antiferromagnets, however, the compensated sublattices are related by crystalline rotations rather than by translation or inversion, producing a momentum-dependent spin splitting that changes sign across the Brillouin zone.

Here we ask what happens when ferromagnetic and altermagnetic orders coexist in the same material. Although both break time-reversal symmetry, they represent fundamentally different forms of magnetic order. In the non-relativistic limit, ferromagnetism corresponds to a momentum-independent spin polarization, whereas altermagnetism is characterized by higher-angular-harmonic spin textures, such as d- or g-wave forms~\cite{SatoPRB2024,krempasky2024altermagnetic,jiang2025metallic,zhang2025crystal}. Their distinct symmetry characters prohibit direct bilinear coupling between the corresponding order parameters. They therefore cannot linearly hybridize, allowing the two orders to remain thermodynamically distinct and, in principle, to undergo separate phase transitions. Higher-order couplings, such as biquadratic terms, are nevertheless allowed and can produce competition or cooperation between their amplitudes.

Spin-orbit coupling (SOC) introduces a qualitatively new interaction between the two orders~\cite{mcclartyLandauTheoryAltermagnetism2024,mazinOriginGossamerFerromagnetism2024,AgetrbergPRL2025,solovyev2026altermagnetism}. While they remain symmetry-distinct order parameters, SOC permits couplings that are odd in each order parameter separately while remaining invariant under time reversal overall. By locking spin orientation to the crystalline lattice, SOC allows the uniform ferromagnetic moment to couple to anisotropic higher-order combinations of the altermagnetic order parameter. The canonical altermagnet $\alpha$-MnTe provides a particularly simple realization of this physics~\cite{mazinOriginGossamerFerromagnetism2024}. Below $T_N\approx$ \SI{307}{K}, MnTe develops an in-plane N\'eel order~\cite{kluczykCoexistenceAnomalousHall2024} (shown on the left side of Fig.~\ref{fig:phasepurity}a) with three equivalent easy-axis orientations and their time-reversal partners, producing six degenerate altermagnetic domains~\cite{xmcd_MnTe,NanodomainNature2024,ZhaoyuMnTe,LiuPRL2026}. This sixfold manifold naturally factorizes into three orientational variants related by 120$^\circ$ rotation, each with two time-reversal partners, forming a $Z_3\times Z_2$ domain structure which is visualized in Fig.~\ref{fig:phasepurity}b. Spin-orbit coupling permits a small uniform out-of-plane magnetization whose sign is locked to the N\'eel vector $\mathbf{L}$, $M_z\propto L^3\sin(3\phi_L)$~\cite{mazinOriginGossamerFerromagnetism2024}. Thus, although the non-relativistic altermagnetic order carries no net moment, SOC provides a symmetry-selective channel connecting a uniform magnetization to its domain structure.

This raises a natural question: what happens if the uniform magnetization is no longer merely a secondary consequence of the altermagnetic order, but becomes an independent thermodynamic order parameter (Fig.~\ref{fig:phasepurity}a right)? We realize this situation in the ferromagnetic altermagnet \mct. Previous studies of powders and thin films have established that Cr-doped MnTe is a ferromagnetic semiconductor with $T_C$ as high as 400~K~\cite{mct_ferromagneticsemicond, mct_nanosheets, mct_p02, polashMagnonSpinTransition2020, magnonbipolar_MCT, kimRoomTemperatureFerromagnetism2025, makovetskiiStructuralMagneticElectrical1997}. Here we show that in bulk single crystals the altermagnetic order persists alongside ferromagnetism up to $x=0.09$, with $T_N$ remaining nearly unchanged while the ferromagnetic ordering temperature is tuned from below to above it. This crossing exposes a pronounced asymmetry between the two orders: when ferromagnetism develops first, the two orders undergo distinct thermodynamic transitions, whereas when altermagnetism develops first, its SOC-mediated coupling to the uniform moment rounds the putative ferromagnetic transition into a crossover. On the other hand, the onset of altermagnetism substantially suppresses the ferromagnetic moment, while the ferromagnetic order acts as a symmetry-selective field on the altermagnetic state, controlling both its anomalous Hall response and its orientation domain populations. Our results show how symmetry-distinct magnetic orders can remain strongly coupled, providing an internal route to controlling altermagnetic domains.

\section{Coexistence of ferromagnetic and altermagnetic order}

\begin{figure*}
    \centering
    \includegraphics[width=0.9\textwidth]{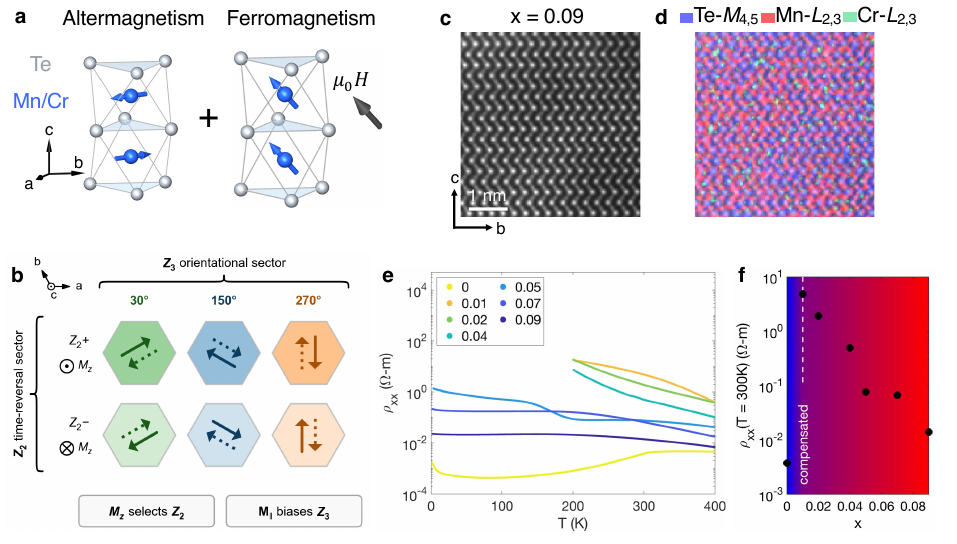}
    \caption{\textbf{Phase schematics and crystal purity.} \textbf{a,} Crystal structure and Mn moment direction of MnTe viewed along the ab-plane (left). Substitution of Cr ions onto the Mn sites leads to ferromagnetic behavior (right). \textbf{b,} Visualization of the $Z_3\times Z_2$ domain structure of MnTe along with the influence of $M$ from the Cr moments. \textbf{c,} HAADF-STEM image of an $x=0.09$ sample showing the same crystal structure as \ch{MnTe}. \textbf{d,} Color map of elemental distribution in the same sample as panel c from atomic-resolution EELS. \textbf{e,} $\rho_{xx}(T)$ for samples with $x$ varying from 0 to 0.09. \textbf{f,} $\rho_{xx}$ at \SI{300}{K} as a function of $x$.}
    \label{fig:phasepurity}
\end{figure*}

\mct\ retains the hexagonal NiAs structure of the parent compound throughout the doping range studied (Fig.~\ref{fig:phasepurity}a). In MnTe, the Mn moments are ferromagnetically aligned within each $ab$ plane and antiferromagnetically stacked along $c$, with an in-plane N\'eel vector $\mathbf{L}$. The two magnetic sublattices are related by the nonsymmorphic $\{C_2|00\tfrac{1}{2}\}$ screw operation that underlies the altermagnetic character of the parent state. Ferromagnetism has previously been reported in Cr-substituted MnTe in polycrystals, nanosheets and thin films, but its microscopic origin remains unsettled~\cite{makovetskiiStructuralMagneticElectrical1997, mct_ferromagneticsemicond, magnonbipolar_MCT, yadakaElectricalResistivityChromium1962, mct_nanosheets, kimRoomTemperatureFerromagnetism2025, yadakaElectricalResistivityChromium1962}. In several proposed scenarios the parent antiferromagnetic state is weakened or replaced by ferromagnetism~\cite{magnonbipolar_MCT, mct_nanosheets, kimRoomTemperatureFerromagnetism2025}. Establishing whether ferromagnetic and altermagnetic orders coexist intrinsically within a homogeneous phase is therefore essential before considering their interaction.

We first establish that the Cr-induced ferromagnetism is an intrinsic bulk property rather than a consequence of magnetic inclusions. Well-faceted single crystals of \mct\ up to $x = 0.10$ were grown using an Sb-flux method previously described for MnTe~\cite{Sbflux_MnTe}; full experimental methods are given in Sec.~\ref{sec:methods}. Powder X-ray diffraction shows that crystals up to $x=0.09$ retain the $\alpha$-MnTe structure (Supplementary Note 1). A minority CrTe phase of approximately 2\% by weight is detectable at the highest doping measured here, but CrTe crystals grown using the same method as the \mct\ crystals are soft ferromagnets ordering below room temperature and therefore cannot account for the hard, in some samples above-room-temperature, ferromagnetism observed here (Supplementary Note 2). Atomic-resolution high-angle annular dark-field imaging scanning transmission electron microscope (HAADF-STEM) and energy-loss spectroscopy (EELS) mapping further show the parent MnTe structure and a homogeneous Cr distribution without detectable Cr-rich aggregates (Fig.~\ref{fig:phasepurity}c,d). Energy-dispersive X-ray spectroscopy (EDX) likewise finds uniform composition across the crystals and no measurable Sb incorporation (Supplementary Note 3). Together, these measurements establish Cr-substituted MnTe as a homogeneous crystallographic phase.

Electrical transport provides an independent bulk probe of the substitution. Remarkably, replacing only 1\% of Mn with Cr raises the room-temperature resistivity by more than three orders of magnitude, after which the resistivity decreases systematically with further Cr substitution (Fig.~\ref{fig:phasepurity}e). This strongly non-monotonic evolution (panel f) is consistent with reaching charge compensation near $x=0.01$ as Cr ions $n$-dope the natively $p$-doped MnTe~\cite{yadakaElectricalResistivityChromium1962, magnonbipolar_MCT}, and is incompatible with a simple mixture of conducting MnTe and magnetic Cr-rich inclusions.

\begin{figure*}
    \centering
    \includegraphics[width=0.9\textwidth]{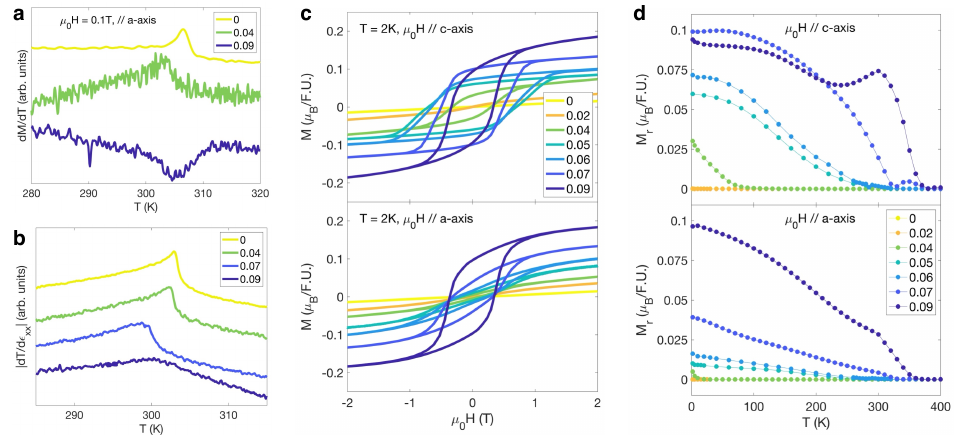}
    \caption{\textbf{Evidence of $T_N$ and permanent magnetization.} \textbf{a,} $\frac{dM}{dT}(T)$ with a \SI{0.1}{T} field applied in-plane for several samples of \mct\ near $T_N$. \textbf{b,} $|\frac{dT}{d\epsilon_{xx}}|(T)$ near $T_N$ for several samples of \mct. \textbf{c,} $M(H)$ at \SI{2}{K} with out-of-plane (top) and in-plane (bottom) magnetic field applied. A hysteresis loop can be observed in Cr-doped samples. \textbf{d,} The zero-field magnetization after field-polarizing the sample ($M_r$) as a function of temperature with out-of-plane (top) and in-plane (bottom) magnetic field applied.}
    \label{fig:phasetransitions}
\end{figure*}

Having established homogeneous substitution, we next determine whether the altermagnetic order survives the emergence of ferromagnetism. Figures~\ref{fig:phasetransitions}a and b track $T_N$ across the doping series using two complementary thermodynamic probes: anomalies in $dM/dT$ (a) and the elastocaloric effect (b), which probes the critical contribution to the heat capacity near a continuous phase transition~\cite{ECE_Ikeda}. Although the anomalies broaden with increasing $x$, likely because of substitutional disorder, their positions remain remarkably stationary: $T_N$ varies by less than \SI{10}{K} across the entire doping range studied here. Furthermore, $d\rho_{xx}/dT$ shows a feature at $T_N$ across the range of $x$ accessed in this study as shown in Supplementary Note 4.

In contrast, the ferromagnetic behavior evolves strongly with Cr concentration. The \SI{2}{K} $M(H)$ curves develop a finite hysteresis for $x\geq0.04$, with a saturated magnetization that increases monotonically with $x$ (Fig.~\ref{fig:phasetransitions}c). The remanent magnetization $M_r$, measured at zero field after field polarization, identifies the onset of ferromagnetic behavior (Fig.~\ref{fig:phasetransitions}d). For $x<0.07$, we denote this onset temperature by $T_{\mathrm{FM}}$, since, as shown below, it represents a crossover rather than a true thermodynamic transition; for $x=0.07$ and 0.09, where the ferromagnetic order develops above $T_N$, we denote the corresponding transition temperature by $T_C$. For $x<0.07$, where $T_{\mathrm{FM}}<T_N$, $M_r$ develops gradually upon cooling, displaying a crossover-like temperature dependence. For $x=0.07$ and 0.09, the out-of-plane $M_r$ rises sharply below $T_C$ but is strongly suppressed on cooling through $T_N$, whereas the in-plane remanence exhibits only a weak kink. This directly reveals a strongly orientation-dependent interaction between the two orders.

\begin{figure}
    \centering
    \includegraphics[width=0.5\textwidth]{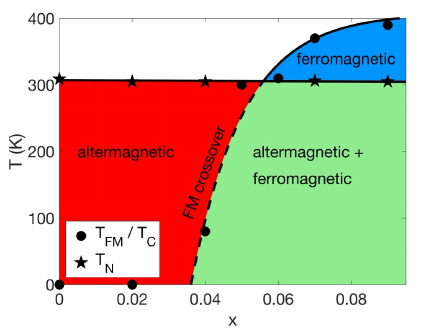}
    \caption{\textbf{$x-T$ phase diagram.} $x-T$ phase diagram of \mct. The coexistence of ferromagnetism and altermagnetism is shown in green.}
    \label{fig:phasediagram}
\end{figure}

Reflective magnetic circular dichroism (RMCD) measurements with a spot size of $\approx$\SI{3}{\micro\m} at \SI{3}{K} provide a local confirmation of the ferromagnetic response: the $x=0.09$ crystal exhibits a much larger hysteretic response than MnTe, without discrete steps indicative of magnetic inclusions on the micron scale (Supplementary Note 4). Together, these measurements produce the $x-T$ phase diagram of Fig.~\ref{fig:phasediagram}. Cr substitution drives the ferromagnetic ordering scale from below to above $T_N$, while $T_N$ itself remains nearly unchanged, creating a broad regime in which ferromagnetic and altermagnetic orders coexist. This phase diagram therefore allows the same two order parameters to be studied in both possible ordering sequences, with ferromagnetism developing either above or below the altermagnetic transition.

When $T_C>T_N$, ferromagnetism first develops through a sharp transition and the altermagnetic order subsequently condenses at a distinct $T_N$. When the ferromagnetic ordering scale lies below $T_N$, however, the situation becomes asymmetric: SOC permits an invariant of the form $\lambda L^3\sin(3\phi_L)M_z,$ so that once $\mathbf{L}$ condenses, the composite altermagnetic order $L^3\sin(3\phi_L)$ acts as a conjugate field on $M_z$, rounding the putative ferromagnetic transition into a crossover.

The two orders also interact through couplings that are even in each order parameter, including the biquadratic term $L^2M_z^2$ and corresponding anisotropic in-plane terms. These naturally account for the strong suppression of the out-of-plane remanence and the much weaker in-plane anomaly at $T_N$. SOC-enabled higher-order couplings are also domain selective. The $L^3\sin(3\phi_L)M_z$ invariant distinguishes the two time-reversal partners within each orientational domain, whereas the anisotropic in-plane coupling distinguishes the three orientational variants of $\mathbf{L}$. We next demonstrate experimentally that the ferromagnetic moment can address both sectors of this $Z_3\times Z_2$ altermagnetic domain manifold.

\section{Altermagnetic domain control with an out-of-plane magnetic moment}

\begin{figure*}
    \centering
    \includegraphics[width=0.9\textwidth]{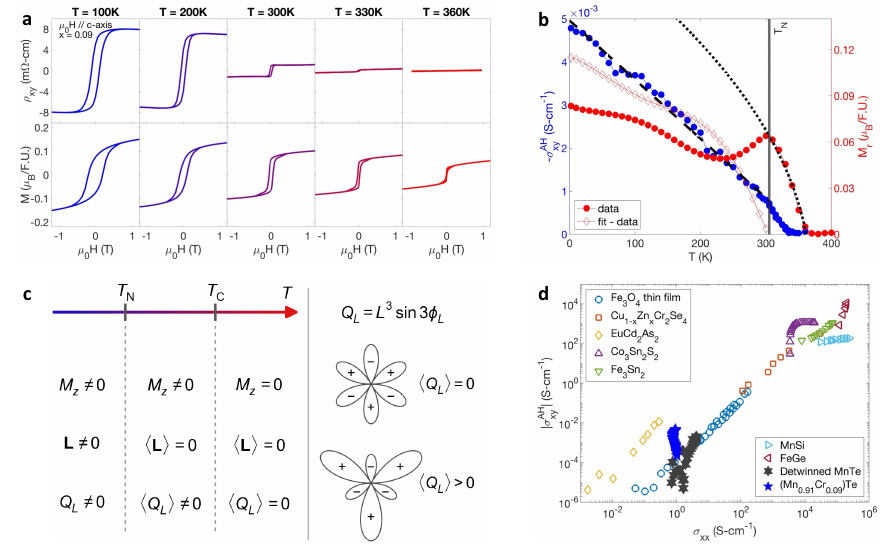}
   \caption{\textbf{Control of the time-reversal domain sector by an out-of-plane magnetization.}
\textbf{a,} Hall resistivity $\rho_{xy}$ (top) and magnetization $M$ (bottom) as a function of out-of-plane magnetic field for an $x=0.09$ sample at several temperatures.
\textbf{b,} Anomalous Hall conductivity $-\sigma_{xy}^{AH}$ and remanent magnetization $M_r$ as a function of temperature, extracted from the measurements in panel a. Dashed lines indicate the temperature dependencies discussed in the main text.
\textbf{c,} The behaviors of $M, L$, and $Q_L$ in different temperature regimes for $x = 0.09$. This is described in detail in the main text.
\textbf{d,} Log--log plot of $|\sigma_{xy}^{AH}|$ as a function of $\sigma_{xx}$ for representative magnetic materials. The $x=0.09$ sample exhibits an anomalous Hall conductivity comparable to that of strain-detwinned MnTe.}
    \label{fig:oopdomaincontrol}
\end{figure*}

The $Z_3\times Z_2$ domain structure introduced above separates the altermagnetic domain manifold into three orientational variants and two time-reversal partners within each orientation. An out-of-plane magnetization couples selectively to the latter sector. In MnTe, the SOC-enabled invariant $\lambda L^3\sin(3\phi_L)M_z$ makes $M_z$ conjugate to the composite altermagnetic order $Q_L=L^3\sin(3\phi_L)$. Reversing $M_z$ therefore switches the bias of $Q_L$, and favors one member of the time-reversal pair, while leaving the three orientational variants degenerate (Fig.~\ref{fig:phasepurity}b). In pure MnTe this coupling can be accessed by an out-of-plane magnetic field acting on the extremely small canted moment~\cite{xmcd_MnTe}. Here we show that in \mct\ the independently ordered ferromagnetic moment provides the same symmetry-breaking field internally, allowing the time-reversal sector of the altermagnetic state to remain selected at zero applied field.

Figure~\ref{fig:oopdomaincontrol}a compares the Hall resistivity $\rho_{xy}$ (top) and magnetization $M$ (bottom) as a function of out-of-plane magnetic field for an $x=0.09$ sample over a wide temperature range. Their temperature dependencies are strikingly different. A clear remanent magnetization persists up to \SI{360}{K}, whereas no anomalous Hall response is resolved at the same temperature. Figure~\ref{fig:oopdomaincontrol}b makes this comparison quantitative. The remanent magnetization onsets at $T_C\approx370$~K and is well described above $T_N$ by the mean-field form $M_r\propto(T_C-T)^{1/2}$ (dotted line), before being suppressed on cooling through the altermagnetic transition at $T_N\approx303$~K. In contrast, $|\sigma_{xy}^{AH}|$ becomes measurable only at temperatures below $\approx$ \SI{330}{K}, despite the already substantial remanent magnetization (50\% of saturated value), and then grows quasi-linearly upon cooling below $T_N$ (dashed line). The anomalous Hall response therefore does not simply track the magnitude of the ferromagnetic moment.

For MnTe, symmetry predicts that the anomalous Hall conductivity is proportional to the composite altermagnetic order $Q_L=L^3\sin(3\phi_L)$~\cite{mcclartyLandauTheoryAltermagnetism2024, mazinAltermagnetismMnTeOrigin2023} In the ordered state, this quantity contains a contribution from the static N\'eel order. For undoped MnTe, neutron diffraction gives $L\propto(T_N-T)^{0.32}$~\cite{criticalexponent_MnTe}, which would imply an approximately linear $L^3$ contribution below $T_N$. The quasi-linear temperature dependence observed in Fig.~\ref{fig:oopdomaincontrol}b is therefore compatible with an altermagnetic origin. We emphasize, however, that we do not measure $L(T)$ directly in Cr-doped MnTe, and there is no reason to assume that its critical behavior is quantitatively identical to that of the parent compound. Moreover, the measured anomalous Hall conductivity does not extrapolate to zero at $T_N$, showing that the static $L^3$ contribution alone cannot account for its temperature dependence.

This behavior naturally follows from the coexistence of ferromagnetic and altermagnetic orders. Because the pre-existing $M_z$ acts as a conjugate field to $Q_L$, the measured expectation value $\langle Q_L\rangle=\langle L^3\sin(3\phi_L)\rangle$ can contain both a contribution associated with static long-range N\'eel order and a field-biased fluctuation contribution. Below $T_N$, both contributions may be present, whereas above $T_N$ the static order vanishes but the fluctuation-induced $\langle Q_L\rangle$ can remain finite. The three orientational variants associated with a given sign of $Q_L$ remain related by threefold rotation, allowing $\langle\mathbf{L}\rangle=0$ while $\langle Q_L\rangle\neq0$. The onset of the anomalous Hall response approximately \SI{25}{K} above $T_N$, together with its smooth evolution through $T_N$, is therefore consistent with a crossover from a fluctuation-dominated response above $T_N$ to a response containing both static and fluctuation contributions below $T_N$. A quantitative separation of these two contributions would require a direct measurement of the altermagnetic order parameter in the Cr-doped crystals. We also note that short-range magnetic correlations have been observed in MnTe $via$ neutron scattering measurements that persist to temperatures well above $T_N$~\cite{baralRealspaceVisualizationShortrange2022}, consistent with this interpretation.

Additional evidence for an altermagnetic contribution to the Hall response is provided by Fig.~\ref{fig:oopdomaincontrol}d, which compares $|\sigma_{xy}^{AH}|$ and $\sigma_{xx}$ for the $x = 0.09$ sample with other magnetic materials. The $x=0.09$ sample reaches an anomalous Hall conductivity comparable to that of mechanically detwinned MnTe~\cite{ZhaoyuMnTe}, supporting a common altermagnetic origin rather than an anomalous Hall response determined solely by the Cr-induced ferromagnetic moment. At the same time, the independently ordered ferromagnetism substantially reduces the field scale required to reverse the trained state compared with pure MnTe~\cite{gonzalezbetancourtSpontaneousAnomalousHall2023,kluczykCoexistenceAnomalousHall2024}. The Cr-induced moment therefore converts the weak magnetic coupling of the parent altermagnet into a robust internal control field for the $Z_2$ time-reversal sector. We next show that an in-plane ferromagnetic moment addresses the complementary $Z_3$ orientational sector.

\section{Control of the orientational domain sector by an in-plane magnetization}

\begin{figure*}
    \centering
    \includegraphics[width=0.9\textwidth]{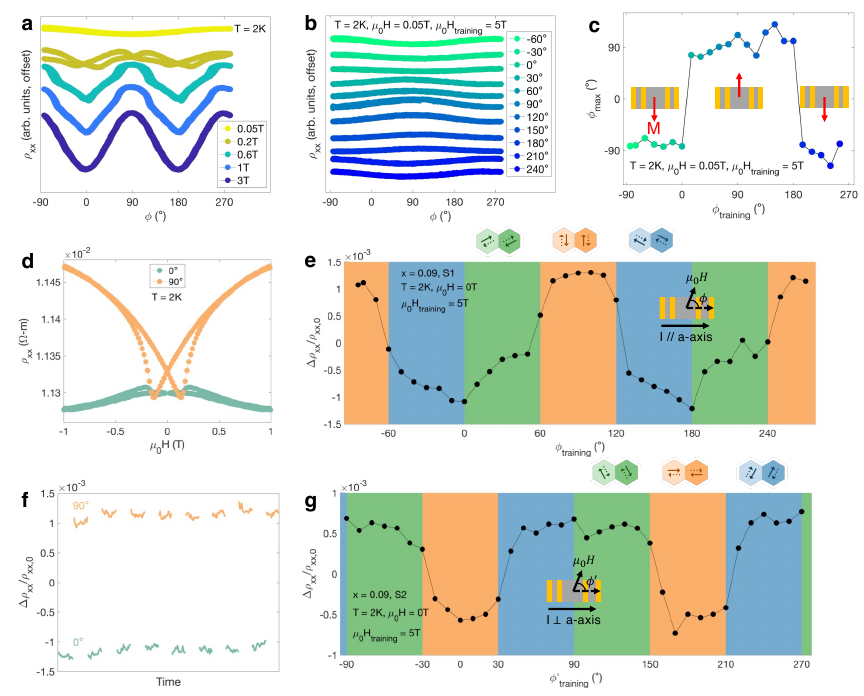}
    \caption{\textbf{Domain control with an in-plane magnetization.} \textbf{a,} $\rho_{xx}(\phi)$ at different $\mu_0H$ at \SI{2}{K} for an $x = 0.09$ sample with current applied parallel to the a-axis (S1). The geometry of the in-plane magnetic field measurements for this sample (used in panels a-f and Fig.~\ref{fig:ipdomaincontrol_summary}) is shown in panel e. \textbf{b,} $\rho_{xx}(\phi)$ at different $\phi_{training}$ with $\mu_0H_{training}=$ \SI{5}{T} at \SI{2}{K} and \SI{0.05}{T}. \textbf{c,} $\phi_{max}$ (defined in main text) as a function of $\phi_{training}$ using the data from panel b. The deduced direction of the magnetization is shown schematically. \textbf{d,} $\rho_{xx}(\mu_0H)$ at 0 and 90$^\circ$. Note that the zero-field resistivity is different depending on magnetic field direction. \textbf{e,} Zero-field $\rho_{xx}$ of S1 as a function of $\phi_{training}$ after applying and removing a \SI{5}{T} magnetic field at $\phi_{training}$ at \SI{2}{K}. Note the features every 60$^\circ$ which are highlighted by color shading. The inferred dominant in-plane altermagnetic domains are shown schematically above the panel with the color matching the shading in the panel. \textbf{f,} Example of reproducibility of the data taken at $\phi_{training}=0^\circ$ and 90$^\circ$ with each segment of data representing \SI{30}{s}. \textbf{g,} The same as panel e but for sample S2 which has the current flowing perpendicular to the a-axis. The geometry of the measurement is shown in the panel.}
    \label{fig:ipdomaincontrol}
\end{figure*}

\begin{figure*}
    \centering
    \includegraphics[width=0.9\textwidth]{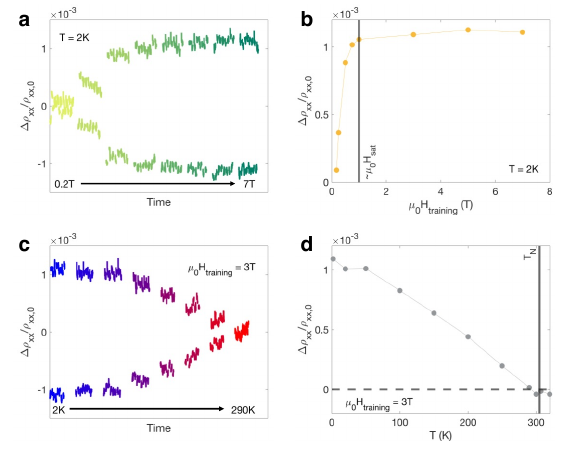}
    \caption{\textbf{Magnetic field and temperature dependence of in-plane magnetization domain control.} \textbf{a,} Similar measurements to those presented Fig.~\ref{fig:ipdomaincontrol}e albeit at various $\mu_0H_{training}$ at \SI{2}{K}. \textbf{b,} The extracted average resistivity $\Delta\rho_{xx}$ between the two $\phi_{training}$ angles as a function of $\mu_0H_{training}$ shown in panel a. The saturation of this curve aligns with the closing of the hysteresis loops from $M(\mu_0H)$ measurements. \textbf{c,} The same measurements at various temperatures at $\mu_0H_{training}=$\SI{3}{T}. A temperature dependent background to $\rho_{xx}$ has been subtracted off. \textbf{d,} The extracted average resistivity $\Delta\rho_{xx}$ between the two $\phi_{training}$ as a function of temperature shown in panel c. This curve reaches the noise floor at roughly $T_N$.}
    \label{fig:ipdomaincontrol_summary}
\end{figure*}

Because the anomalous Hall response depends on $\sin(3\phi_L)$, it is identical for the three orientational variants and thus cannot distinguish the $Z_3$ sector. We therefore use the longitudinal resistivity as the method for domain readout in an $x = 0.09$ sample with the transport geometry shown in the top right of Fig.~\ref{fig:ipdomaincontrol}e. We first establish the behavior of the in-plane ferromagnetic moment in our transport geometry by rotating the sample in a fixed field (panel a). At fields well above the saturation field (which is $\approx$ \SI{1}{T}), the angle-dependent magnetoresistance (AMR) displays the conventional two-fold response expected when $\mathbf{M}$ follows the applied field, whereas pronounced hysteresis appears at intermediate fields as the magnetization switches. Well below the coercive field, the angular response becomes one-fold as seen in the \SI{0.05}{T} data. In this low-field regime, the magnetization is pinned by magnetic anisotropy and the maximum angle of the one-fold AMR provides a readout of the orientation of the pinned magnetization, as established in previous angular magnetotransport studies~\cite{heronElectricFieldInducedMagnetizationReversal2011,mibuReversibleMagnetizationProcess1996}. This is investigated by systematically applying a \SI{5}{T} magnetic field and then reducing this field to \SI{0.05}{T} at different $\phi_{training}$, then performing a full angle sweep as shown in Fig.~\ref{fig:ipdomaincontrol}b. After removal of the training field, the remanent magnetization relaxes to one of two orientations, $\mathbf{M}\parallel\pm90^\circ$ relative to the current, independent of the training angle $\phi_{\mathrm{training}}$. This can clearly be seen in panel c where $\phi_{max}$ (the angle at which $\rho_{xx}$ is maximum $\pm$360$^\circ$) is plotted as a function of $\phi_{training}$. The two states observed differ only by $\mathbf{M}\rightarrow-\mathbf{M}$ and therefore have the same conventional ferromagnetic AMR. Thus, although $\mathbf{M}$ follows $\phi_{\mathrm{training}}$ during high-field training, conventional ferromagnetic AMR cannot account for a systematic dependence of the final zero-field resistivity on the training-field direction.

Despite this fact, magnetic field sweeps at different angles $\phi_{training}$ display different zero-field resistivites, demonstrating that the final zero-field state retains a memory of the direction of the training field. Fig.~\ref{fig:ipdomaincontrol}d displays a clear observation of this effect in the 0 and $90^\circ$ data. Despite the small change of the resistivity, approximately $0.2\%$, switching between trained states is highly reproducible. Panel f shows the reproducibility of the data repeatedly switching between $\phi_{training}=0^\circ$ and 90$^\circ$ with \SI{30}{s} dwell time between switching. Because the remanent ferromagnetic moment has the same orientation axis relative to the current in each trained state, this angular dependence cannot arise from conventional AMR of $\mathbf{M}$ and instead reflects a training-induced redistribution of the altermagnetic orientation domains.

Panel e extends this by presenting the zero-field $\rho_{xx}$ measured after applying and removing a \SI{5}{T} in-plane field at many different $\phi_{\mathrm{training}}$. The resistivity exhibits a dominant two-fold modulation with superimposed features at $60^\circ$ intervals highlighted by different colors. The three orientational variants of MnTe have N\'eel vectors along the three equivalent $\langle1\overline{1}0\rangle$ axes. For a Hall bar with current along $\mathbf{a}$, the domain with $\mathbf{L}\parallel\mathbf{j}$ produces a different longitudinal resistivity from the two symmetry-related domains with $\mathbf{L}$ rotated by $\pm60^\circ$. The three-domain manifold therefore projects onto two longitudinal-resistance values, producing the dominant two-fold modulation. As the training magnetization is rotated, the anisotropic coupling successively favors different members of this discrete set, producing the additional features at $60^\circ$ intervals. Importantly, these features reflect selection among the three orientational variants and do not distinguish the two time-reversal partners within each orientation, consistent with the even-in-$\mathbf{L}$ form of the in-plane coupling. The schematics above panel e illustrate the inferred dominant orientational domain population for representative training directions with the majority domain's color matching the shading in the panel. 

To further confirm these observations, a sample with current perpendicular to the $a$-axis (S2) is also measured. As shown in panel g, despite the same ferromagnetic moment directions, the $60^\circ$ features have been shifted by $30^\circ$, with two high-resistivity sections and one low-resistivity section. These results are consistent with the $90^\circ$ rotation of the $a$-axis with respect to current, and further confirm that the zero-field resistivity is determined by the altermagnetic orientational domains. More characterization and results of another sample can be found in Supplementary Note 7.

Returning to S1, the zero-field resistivity difference \(\Delta\rho_{xx}\) between states trained at \(0^\circ\) and \(90^\circ\) rises rapidly with training field and saturates on the field scale at which the ferromagnetic hysteresis loop closes (Figure~\ref{fig:ipdomaincontrol_summary}a,b). Once \(\mathbf M\) is fully polarized, stronger training fields produce no further increase in \(\Delta\rho_{xx}\), indicating that domain writing is mediated by the ferromagnetic moment rather than directly by the applied field.

The temperature dependence provides the complementary connection to the altermagnetic order. As shown in Fig.~\ref{fig:ipdomaincontrol_summary}c and d, $\Delta\rho_{xx}$ decreases upon warming and falls into the noise floor near $T_N$. Taken together, the field and temperature dependencies identify the two order parameters involved in the switching: the state is written by the ferromagnetic moment during field training but stored in the altermagnetic orientation-domain population after the field is removed. This nonvolatile switching contrasts with the reversible in-plane-field-induced orientational alignment recently observed by neutron diffraction in undoped $\alpha$-MnTe~\cite{pan2026detwinning}.

\section{Discussion}


Cr-induced ferromagnetism provides internal symmetry-breaking fields that address both sectors of the altermagnetic domain manifold: its out-of-plane component selects the \(Z_2\) time-reversal sector, while its in-plane component biases the \(Z_3\) orientational sector. The trained domain state remains stable at zero applied field, establishing an internal route to control the factorized \(Z_3\times Z_2\) manifold. From the broader perspective of intertwined orders, the coupling between ferromagnetism and altermagnetism is reminiscent of fluctuation-driven composite order in iron-based superconductors. In the spin-driven picture of nematicity, the nematic order parameter, $\phi_{\mathrm{nem}}\sim|\boldsymbol{\Delta}_X|^2-|\boldsymbol{\Delta}_Y|^2$, is a composite of the underlying stripe antiferromagnetic order and can develop before long-range magnetic order~\cite{Fernandes2014}. In MnTe, the corresponding SOC-active composite is $Q_L=L^3\sin(3\phi_L)$, which transforms in the same symmetry channel as $M_z$. The analogy is not exact: in Cr-doped MnTe the ferromagnetic moment is an independent order parameter whose bare instability is controlled by Cr concentration, rather than a purely vestigial order of the altermagnet. Nevertheless, fluctuations of $Q_L$ above $T_N$ renormalize the ferromagnetic susceptibility through the coupling $-\lambda M_zQ_L$. Therefore, it is likely that the growing altermagnetic correlations may enhance the underlying Cr-induced ferromagnetic instability and contribute to the occurrence of $T_C>T_N$. Once $M_z$ condenses, the same coupling acts in the opposite direction: $M_z$ becomes a conjugate field to $Q_L$ and biases a finite $\langle Q_L\rangle$ even while $\langle\mathbf{L}\rangle=0$, providing a natural mechanism for the anomalous Hall response observed above $T_N$. Below $T_N$, this cooperative composite coupling coexists with the competing biquadratic interaction $L^2M_z^2$, which accounts for the suppression of the uniform moment when static altermagnetic order develops.


Recent theory also suggests that orbital magnetization may play an important role in separating the ferromagnetic and Hall responses. First-principles calculations for MnTe find an SOC-induced orbital magnetization substantially larger than the corresponding spin magnetization~\cite{YePRL2026}, while recent work relates the intrinsic anomalous Hall conductivity of altermagnets directly to the orbital magnetization through a generalized St\v{r}eda relation~\cite{zhao2026residualorbitalmagnetizationgoverns}. These results suggest that the Cr-induced spin magnetization and the Hall-active orbital response need not scale together, providing a possible microscopic basis for the pronounced difference between $M(T)$ and $\sigma_{xy}^{AH}(T)$ observed here. We emphasize, however, that the high-temperature ferromagnetism in the $x=0.09$ sample already develops above $T_N$ and therefore cannot originate solely from orbital magnetization associated with static altermagnetic order. Element- and orbital-resolved XMCD across both $T_C$ and $T_N$, together with direct measurements of $L(T)$ by neutron scattering, would provide a direct test of this picture and allow the static and fluctuation contributions to the coupled ferromagnetic--altermagnetic state to be separated.

\section{Methods}
\label{sec:methods}
\subsection{Crystal Growth}
Single crystals of \mct\ were grown as following the recipe for MnTe presented in Ref.~\cite{Sbflux_MnTe}: Mn chunks (99.98\%, cleaned by being heated at 800$^\circ$C for about 12 hours in flowing Ar/\ch{H2} gas (5\% molar concentration of hydrogen) and then stored in an Ar glovebox until use), Cr powder (99.99\%), Te lump (99.999+\%), and Sb shot (99.999\%) were combined in atomic ratios of Mn:Cr:Te:Sb 1-$x_{nom}$:$x_{nom}$:1.2:5. These mixtures were loaded in Canfield crucible sets~\cite{Canfield2016} and sealed under vacuum in quartz tubes. They were then heated to 1000$^\circ$C in 5 hours, held at this temperature for 12 hours, then cooled to 650$^\circ$C in 70 hours. Then the growths were decanted in a centrifuge to separate the excess flux from the crystals. For $x_{nom} \leq 0.1$ well-faceted single crystals of \mct\ with sides longer than \SI{1}{mm} were formed, whereas higher $x_{nom}$ provided crystals smaller in size and often containing physical holes with a maximum $x_{EDX}$ of about 0.10.

\subsection{Chemical and structural charcterization}
The doping concentration \textit{x} of several crystals from each batch was determined from EDX with a Sirion XL30 scanning electron microscope. Measurements were taken at at least 8 regions spread out over at least 1 mm$^2$ area of the sample. Each crystal measured from a given batch had nearly the same $x_{EDX}$. Supplementary Note 3 contains further details of the EDX measurements.

Powder X-ray diffraction measurements were performed on crushed single crystals using the Rigaku MiniFlex 600 system, with a Cu source and Hy-Pix 400MF 2D-detector.

\subsection{Magnetization and transport measurements}
Magnetization measurements were performed with the vibrating sample magnetometer option of either a Quantum Design Magnetic Property Measurement System 3 or a Quantum Design Dynacool Physical Property Measurement System (PPMS). Samples were attached to a quartz paddle with GE varnish such that the magnetic field was either out-of-plane (parallel to the c-axis) or in-plane.

Transport measurements were performed on samples that were polished and cut by a wire saw to be bars with dimensions roughly \SI{1}{mm} $\times$ \SI{0.5}{mm} $\times$ \SI{0.1}{mm}. Gold pads were sputtered onto the sample to make 6 point (Hall pattern) contacts. Silver epoxy and gold wires were used to contact the samples. These measurements were performed in a PPMS with standard lock-in techniques. To eliminate any contributions from contact misalignment, the Hall resistivity was anti-symmetrized with magnetic field. Samples with $0.01\leq x \leq 0.06$ were highly insulating upon cooling down which hindered reliable measurements at low temperatures.

\subsection{Elasto-caloric measurements}
Uniaxial stress measurements were performed by gluing a sample across the gap of a home-made three piezoelectric strain cell, modeled after the commercial Razorbill CS-100 strain cell. The sample was secured to the titanium cell across a gap of \SI{0.5}{mm} using Stycast 2850FT Epoxy. For elastocaloric measurements, an AC voltage of \SI{5}{V} root mean square was applied to the inner piezoelectric stack of the strain cell, which corresponded to creating an AC displacement of the sample of approximately 0.01\% of its length. The temperature fluctuations of the sample induced by this AC strain were measured using a home-made Type E thermocouple comprised of chromel and constantan wires that were silver pasted to each other and the sample. The AC frequency was experimentally determined by measuring the elastocaloric signal at \SI{300}{K} for frequencies in the range of 1-\SI{100}{Hz}, and choosing the frequency with the largest response. This frequency was at the peak of the relevant thermal transfer function, which did not observably shift in the temperature range measured~\cite{ECE_Ikeda}.

\subsection{Transmission electron microscope measurements}
Transmission electron microscope (TEM) lamellae were prepared from \mct\ crystals using a ThermoFisher Scios 2 DualBeam focused ion beam system, with a final polishing process at \SI{2}{kV}. A \SI{2}{\micro\m} thick platinum or carbon protective layer was deposited on the top of the lamellae. All TEM lamellae were further thinned down to an electron-transparent thickness utilizing a Fischione’s Model 1040 NanoMill TEM specimen preparation system at an operating voltage of \SI{900}{V} and a beam current of \SI{140}{pA}.

Atomic resolution $Z$-contrast HAADF-STEM imaging was carried out on a Nion UltraSTEM\textsuperscript{TM} 100 microscope, equipped with a spherical aberration corrector, operating at \SI{60}{kV}. The HAADF images were acquired with a convergence semi-angle of \SI{32}{mrad} and collection semi-angles from 80 to \SI{200}{mrad}. To minimize sample drift effects, we recorded each STEM dataset as 20 successive image frames using a fast scan (\SI{1}{\micro\s}-pixel$^{-1}$ dwell time), followed by post-processing
drift correction through image registration.

The core-loss electron energy loss spectra were acquired using a Nion Iris spectrometer and Dectris ELA high-speed electron-counting detector attached to the Nion UltraSTEM\textsuperscript{TM} at Oak Ridge National Laboratory. Spectrum images were acquired using a pixel dwell time of \SI{10}{ms}, an energy dispersion of \SI{0.9}{eV} per channel, and a collection semi-angle of \SI{40}{mrad}. The scanned area was 6 $\times$ \SI{6}{nm^2} with a pixel size of \SI{0.027}{nm}. The spectrum images were denoised using PCA by selecting the first 20 components according to the scree plots. Cr $L_{2,3}$, Te $M_{4,5}$, and Mn $L_{2,3}$ edges were used for elemental mapping.

\subsection{RMCD measurements}
RMCD measurements were performed on bulk single crystals in a Quantum Design OptiCool cryostat equipped with a superconducting magnet providing out-of-plane fields of up to \SI{7}{T}. A \SI{532}{nm} laser with an incident power of \SI{100}{\micro\W} was focused to a $\approx$\SI{3}{\micro\m} spot at near-normal incidence on the sample surface.

The incident beam was linearly polarized at 45$^\circ$ relative to the optical axes of a photoelastic modulator (PEM). The PEM was operated at $f_{PEM}$ = \SI{50}{kHz} with a quarter- wave peak retardance to introduce a sinusoidally varying phase difference between orthogonal electric-field components, producing opposite circular polarization states at the positive and negative retardance extrema. The laser intensity was additionally modulated at $f_{chopper}\approx$ \SI{4}{kHz} using an optical chopper to provide a reference signal proportional to the reflected intensity.

The reflected light was detected using a silicon photodetector whose output was demodulated at $f_{PEM}$ and $f_{chopper}$ using a Zurich Instruments HF2LI lock-in amplifier. The resulting signals, $I_{PEM}$ and $I_{chopper}$, were used to calculate the reported RMCD signal as RMCD~$=\frac{I_{PEM}}{J_1(\pi/2)I_{chopper}}$, where $J_1$ is the first-order Bessel function.

\subsection{Free energy and symmetry-allowed couplings}

We describe the coupled orders by the in-plane N\'{e}el vector $\mathbf{L} = L(\cos\phi_L, \sin\phi_L)$ and the magnetization $\mathbf{M} = (\mathbf{M}_\parallel, M_z)$. The screw symmetry $\{C_2\,|\,00\tfrac{1}{2}\}$ of the parent lattice reverses $\mathbf{L}$ while leaving $\mathbf{M}$ invariant, so all terms in the free energy must be even in $\mathbf{L}$ except where $\mathbf{L}$ enters through combinations that are themselves screw-even. To the order relevant here,
\begin{align}
F &= F_L + F_M + F_{\mathrm{int}}, \nonumber\\
F_L &= \frac{a_L}{2}\,(T - T_N^0)\,L^2 + \frac{b_L}{4}\,L^4
      - \frac{v}{6}\,L^6\cos 6\phi_L, \nonumber\\
F_M &= \frac{a_M}{2}\,(T - T_C^0)\,|\mathbf{M}|^2
      + \frac{b_M}{4}\,|\mathbf{M}|^4
      + \frac{K}{2}\,M_z^2, \nonumber\\
F_{\mathrm{int}} &= \frac{\beta}{2}\,L^2 M_z^2
  + \frac{\bar{\alpha}}{2}\,L^2\,|\mathbf{M}_\parallel|^2
  + \frac{g}{2}\Big[(L_x^2 - L_y^2)(M_x^2 - M_y^2) \nonumber\\
  &\quad + 4\,L_x L_y\, M_x M_y\Big]
  - \lambda\,L^3 \sin(3\phi_L)\,M_z, \label{eq:GL}
\end{align}
The bilinear invariant $\mathbf{L}\cdot\mathbf{M}$ is forbidden by the screw symmetry; this is the symmetry protection that permits the two time-reversal-breaking orders to coexist as independent thermodynamic phases. The leading allowed interactions are the biquadratic terms ($\alpha$, $\beta$, $\gamma$) and the cubic invariant $\lambda$, which requires spin-orbit coupling at third order~\cite{mazinOriginGossamerFerromagnetism2024, mcclartyLandauTheoryAltermagnetism2024} and is responsible for the gossamer moment of pure MnTe. The sixfold anisotropy $v$ discretizes $\phi_L$ to six domain orientations. In this note we work at mean-field level and address two observations reported in the main text: the orientation-selective suppression of the remanent magnetization below $T_N$, and the insensitivity of $T_N$ to Cr doping. Both are controlled by the single coupling $\beta$.

\subsection{Suppression of the out-of-plane remanence below $T_N$}

Consider a sample with $T_C > T_N$ trained along the $c$-axis, so that $M_z \neq 0$, $\mathbf{M}_\parallel = 0$ above $T_N$. Minimizing $F_M$ with the biquadratic term included, the out-of-plane magnetization below $T_N$ satisfies
\begin{equation}
\left[a_M (T - T_C^0) + K \right] M_z + b_M M_z^3
  + \beta\, L^2(T)\, M_z = 0,
\end{equation}
so the condensation of $\mathbf{L}$ acts as a temperature-dependent positive shift of the quadratic coefficient of $M_z$. To leading order the fractional suppression of the equilibrium magnetization is
\begin{equation}
\frac{\delta (M_z^2)}{M_z^2}
  = -\,\frac{\beta\, L^2(T)}{a_M\,(T_C - T) - K}
  \;\simeq\; -\,\frac{\beta\, L^2(T)}{a_M\,(T_C - T)},
\label{eq:suppression}
\end{equation}
valid for $T$ not too close to $T_C$. Because $L^2(T)$ grows continuously below $T_N$, the suppression onsets at $T_N$ and deepens on cooling, competing with the growth of the uncoupled magnetization as $T_C - T$ increases; the net result is the non-monotonic $M_r(T)$ observed for the $x = 0.07$ and $0.09$ samples. Using the neutron-diffraction result $L \propto (T_N - T)^{0.32}$~\cite{criticalexponent_MnTe}, the predicted onset form of the suppression is $(T_N - T)^{0.64}$.

For in-plane training, the suppression of $M_r$ at $T_N$ is governed by the effective coupling $\beta_\parallel^{\mathrm{eff}} = \tilde{\alpha} + g\,\overline{\cos 2(\phi_L - \phi_M)}$, where the overline denotes an average over the altermagnetic domain population; the quadrupolar channel contributes with full weight only in a domain-selected state and partially cancels in a multi-domain state. Experimentally, the in-plane $M_r(T)$ curves show a kink at $T_N$ that is distinctly weaker than the out-of-plane suppression. Within mean field the ratio of the two anomalies directly measures $\beta_\parallel^{\mathrm{eff}}/\beta$. The strong inequality $\beta_\parallel^{\mathrm{eff}} < \beta$ is consistent with the easy-plane magnetic anisotropy of the parent compound, in which out-of-plane spin components are energetically costly and hence most sensitive to the exchange reconstruction at $T_N$. We emphasize that all three couplings are independent symmetry-allowed constants; no symmetry requires the in-plane anomaly to vanish, and its finite value, together with the finite $g$ required for the in-plane domain selection demonstrated in the main text, confirms that the coupling operates in both channels.

\subsection{Quantitative estimate of $\beta$}

The coefficients of $F_M$ are fixed by the measured ferromagnetic properties of the $x = 0.09$ sample. From the out-of-plane $M_r(T)$ curve we read off the suppression at its deepest point near $T \approx 270$~K: $M_r$ is reduced from an interpolated uncoupled value of $\approx 0.09\,\mu_B$/F.U. to $\approx 0.07\,\mu_B$/F.U., giving $\delta(M_z^2)/M_z^2 \approx 0.4$. The N\'{e}el order parameter at this temperature, normalized to the saturated Mn moment $L_{\mathrm{sat}} \approx 5\,\mu_B$/Mn, is $L^2(270\,\mathrm{K}) \approx (1 - 270/300)^{0.64} \times L_{\mathrm{sat}}^2 \approx 6\,\mu_B^2$. With $T_C - T \approx 80$~K, Eq.~(\ref{eq:suppression}) yields
\begin{equation}
\beta \approx 0.4 \times \frac{a_M \times 80\,\mathrm{K}}
{6\,\mu_B^2} \approx 5\, a_M\,\mathrm{K}/\mu_B^2.
\end{equation}

\subsection{Back-action on $T_N$}

The same coupling shifts the altermagnetic transition. With $M_z$ condensed above $T_N$, the quadratic coefficient of $L$ becomes $a_L (T - T_N^0) + \beta M_z^2$, so
\begin{equation}
\Delta T_N = -\,\frac{\beta\, M_z^2(T_N)}{a_L}
  = -\,\frac{a_M}{a_L}\times 5\,\mathrm{K}/\mu_B^2 \times
  M_z^2(T_N).
\end{equation}
With $M_z(T_N) \approx 0.08\,\mu_B$/F.U., i.e. $M_z^2 \approx 6\times 10^{-3}\,\mu_B^2$, this gives $|\Delta T_N| \approx 0.03\,\mathrm{K} \times (a_M/a_L)$. The ratio $a_M/a_L$ of the inverse Curie constants of the two orders is of order unity to ten for comparable moment densities, so the predicted shift is of order tens of millikelvin. The physical origin of this hierarchy is the extreme asymmetry of the two order parameters: the suppression of $M_z$ scales with $L^2 \sim 6\,\mu_B^2$, while the back-action on $T_N$ scales with $M_z^2 \sim 6\times 10^{-3}\,\mu_B^2$, three orders of magnitude smaller. The same coupling that produces a clearly visible ($\sim$20\%) suppression of the remanent magnetization is therefore quantitatively guaranteed to leave $T_N$ unresolved within our experimental sensitivity of a few kelvin. The observed insensitivity of $T_N$ to Cr doping is not evidence of decoupled orders; it is the mean-field consequence of a coupling whose magnitude is independently fixed by the magnetization data.

\subsection{Role of the cubic invariant}

The cubic term $\lambda L^3 \sin(3\phi_L) M_z$ is linear in $M_z$ and therefore cannot suppress the magnitude of the remanence; its equilibrium contribution to $M_z$ in pure MnTe is the gossamer moment of order $10^{-5}\,\mu_B$/Mn~\cite{mazinOriginGossamerFerromagnetism2024, kluczykCoexistenceAnomalousHall2024}, four orders of magnitude below the scale of Eq.~(\ref{eq:suppression}). Its role in the doped system is qualitatively different: with $M_z$ supplied independently by the Cr order, the term acts as a field of strength $\lambda M_z$ conjugate to the sign of $L^3\sin(3\phi_L)$, lifting the degeneracy between the two time-reversal partners within each of the three orientation domains. We note that the value of $\lambda$ in the doped system is not constrained by the gossamer moment of pure MnTe, since Cr substitution alters both the magnitude of $M_z$ entering the selection energy and, through the shift of the chemical potential, potentially the coefficient $\lambda$ itself. Together with the in-plane biquadratic term $\gamma$, which biases the orientation of $\mathbf{L}$ when $\mathbf{M}_\parallel$ is trained in-plane, the cubic term provides the two-channel domain selection demonstrated in the main text: an out-of-plane moment reduces the sixfold degeneracy to three (time-reversal selection), and an in-plane moment reduces it further (orientation selection).

\section*{Acknowledgments}

\subsection*{Funding}

This work was primarily supported by the University of Washington Molecular Engineering Materials Center, a U.S. National Science Foundation Materials Research Science and Engineering Center (DMR-2308979). The work done by Y.X. was supported in part by the Gordon and Betty Moore Foundation, Grant GBMF13842. Part of this work was conducted at the Molecular Analysis Facility, a National Nanotechnology Coordinated Infrastructure site at the University of Washington which is supported in part by the National Science Foundation (grant NNCI-1542101), the University of Washington, the Molecular Engineering \& Sciences Institute, the Clean Energy Institute, and the National Institutes of Health. 

\subsection*{Author contributions}

J.M.D. and Y.X. grew the crystals and performed the magnetotransport and thermodynamic measurements. G.R. performed the STEM experiments and analyzed the data with guidance from J.C.I.. M.N. and Y.S. performed the RMCD measurements with guidance from X.X.. J.-H.C. oversaw the project. All authors contributed to the writing of the manuscript.

\section*{Competing interests}
The authors declare no competing interests.

\section*{Data availability}

All data needed to evaluate the conclusions in the paper are present in the paper and/or the Supplementary Information. 

\newpage

\bibliography{main}

\clearpage

\setcounter{MaxMatrixCols}{10}
\graphicspath{{./}{./figs/}}

\renewcommand{\thefigure}{S\arabic{figure}}
\setcounter{figure}{0}
\renewcommand{\thetable}{S\arabic{table}}
\setcounter{table}{0}
\renewcommand{\theequation}{S\arabic{equation}}
\setcounter{equation}{0}

\captionsetup{justification   = raggedright, singlelinecheck = false}

\begin{center}
\section*{Supplementary Materials for ``Intertwined ferromagnetic and altermagnetic orders in Cr-doped MnTe"}
\end{center}

\subsection*{Supplementary Note 1: Powder X-ray Diffraction}

\begin{figure*}[h]
    \centering
    \includegraphics[width=0.9\textwidth]{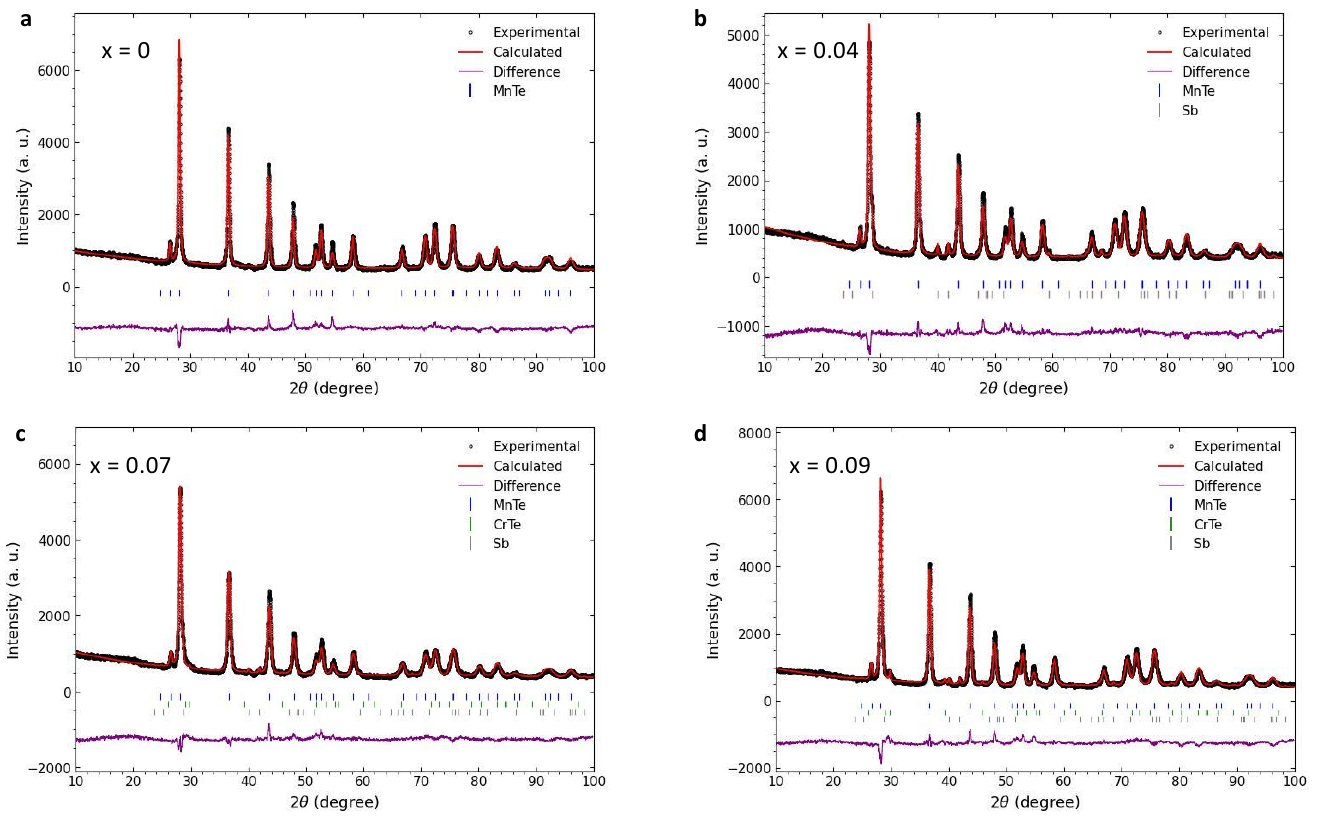}
    \caption{\textbf{PXRD on \mct.} \textbf{a, b, c, d, } PXRD spectra and fits for samples with $x = 0$, 0.04, 0.07, and 0.09, respectively.}
    \label{fig:pxrd}
\end{figure*}

Fig.~\ref{fig:pxrd} presents the powder X-ray diffraction (PXRD) spectra and fits for several samples of \mct. The fits were conducted by including three phases: MnTe with the lattice parameters allowed to vary due to the chemical pressure induced by Cr doping, Sb (due to residual flux on the surfaces of the crystals), and (isostructural to MnTe) CrTe which seems to grow in small amounts in \mct. The relative weight percentages of these phases are outlined in Table~\ref{table:weightpercents} and the extracted lattice parameters of the \mct\ are listed in Table~\ref{table:latticeparameters}. With increasing $x$ the lattice parameters decrease which is consistent with literature~\cite{magnonbipolar_MCT,mct_ferromagneticsemicond}.

\begin{table}[h]
\begin{tabular}{|l|l|l|l|}
\hline
$x$ & \mct\ (\%) & Sb (\%) & CrTe (\%) \\ \hline
0  & 100  & - & - \\ \hline
0.04  & 95.35 & 4.65 & - \\ \hline
0.07  & 97.24 & 2.16 & 0.60 \\ \hline
0.09  & 95.15 & 2.64 & 2.21 \\ \hline
\end{tabular}
\caption{\textbf{Weight percentages of each phase from PXRD.}}
\label{table:weightpercents}
\end{table}

\begin{table}[h]
\begin{tabular}{|l|l|l|}
\hline
$x$ & a (\AA) & c (\AA) \\ \hline
0  & $4.1496 \pm 0.0002$ & $6.7134 \pm 0.0004$ \\ \hline
0.04  & $4.1465 \pm 0.0002$ & $6.7049 \pm 0.0004$ \\ \hline
0.07  & $4.1464 \pm 0.0002$ & $6.7048 \pm 0.0004$ \\ \hline
0.09  & $4.1412 \pm 0.0001$ & $6.7018 \pm 0.0002$ \\ \hline
\end{tabular}
\caption{\textbf{\mct\ lattice parameters from PXRD.}}
\label{table:latticeparameters}
\end{table}

\subsection*{Supplementary Note 2: Characterization of CrTe}

\begin{figure*}
    \centering
    \includegraphics[width=0.9\textwidth]{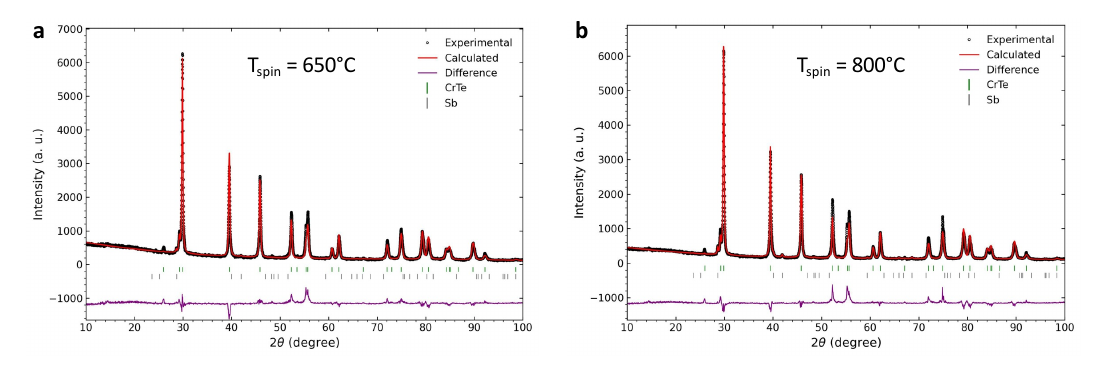}
    \caption{\textbf{PXRD on CrTe.} \textbf{a, b,} PXRD spectra for the CrTe spun at 650 and 800$^\circ$C, respectively. Both spectra can be well fit to the expected CrTe diffraction pattern with a small amount of residual Sb flux.}
    \label{fig:CrTe_PXRD}
\end{figure*}

\begin{figure*}
    \centering
    \includegraphics[width=0.9\textwidth]{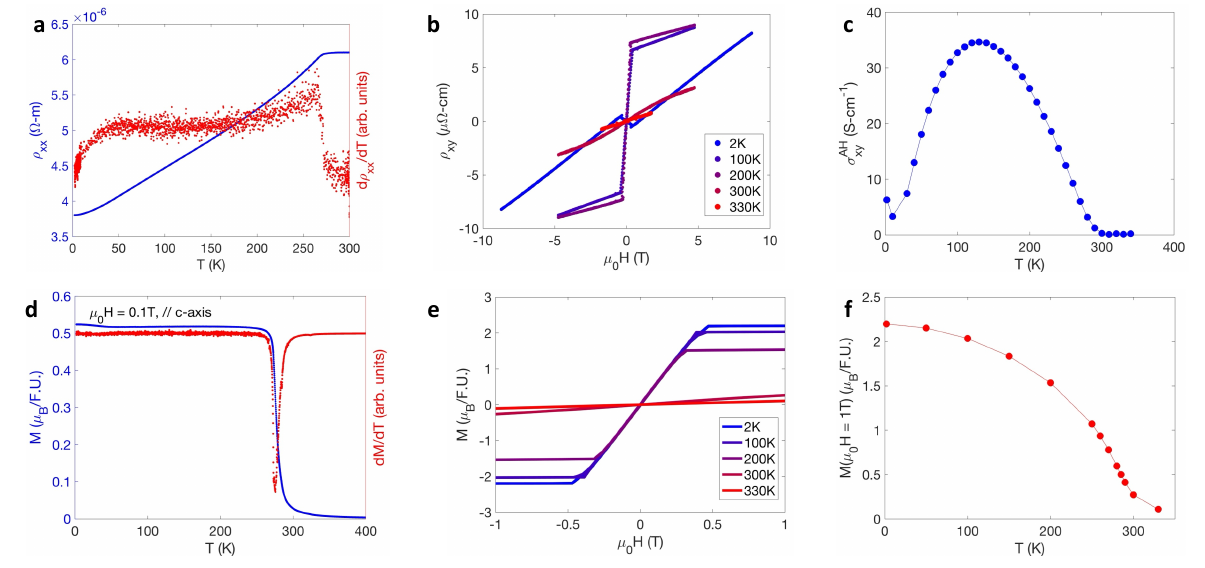}
    \caption{Supplementary Figure 3. \textbf{Characterization of CrTe spun at 650$^\circ$C.} \textbf{a,} $\rho_{xx}(T)$ (blue) and $\frac{d\rho_{xx}(T)}{dT}$ (red). \textbf{b,} $\rho_{xy}(\mu_0H)$ at several temperatures. \textbf{c,} $\sigma_{xy}^{AH}(T)$. \textbf{d,} $M(T)$ (blue) and $\frac{dM(T)}{dT}$ (red) with a \SI{0.1}{T} magnetic field along the c-axis. \textbf{e,} $M(\mu_0H)$ at several temperatures with the same magnetic field direction. \textbf{f,} $M(T)$ at \SI{1}{T} from the $M(\mu_0H)$ measurements.}
    \label{fig:CrTe650}
\end{figure*}

\begin{figure*}
    \centering
    \includegraphics[width=0.9\textwidth]{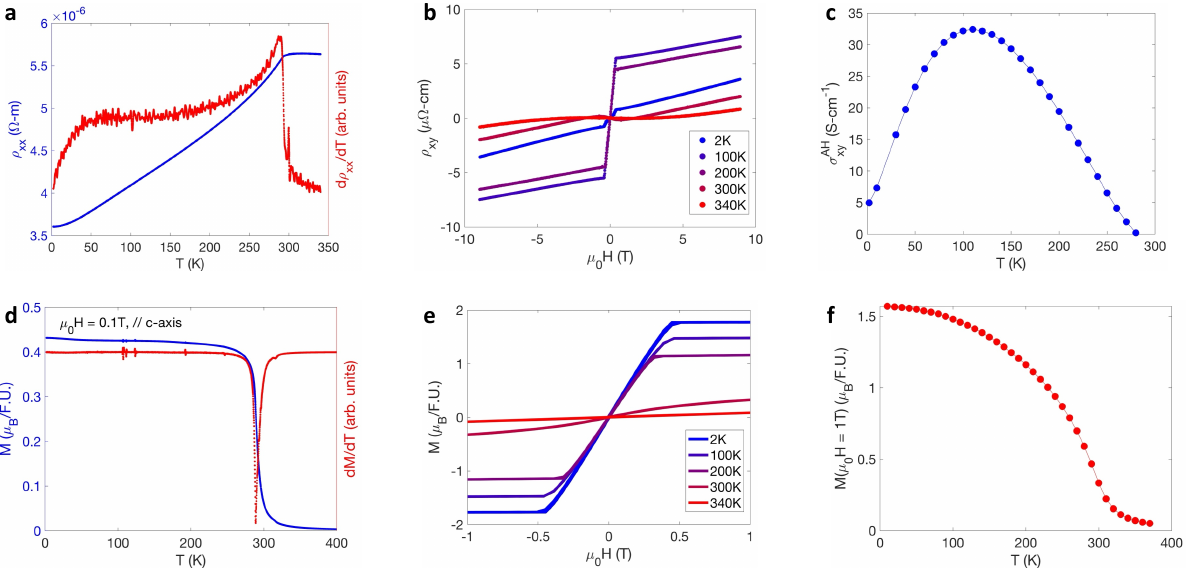}
    \caption{Supplementary Figure 4. \textbf{Characterization of CrTe spun at 800$^\circ$C.} \textbf{a,} $\rho_{xx}(T)$ (blue) and $\frac{d\rho_{xx}(T)}{dT}$ (red). \textbf{b,} $\rho_{xy}(\mu_0H)$ at several temperatures. \textbf{c,} $\sigma_{xy}^{AH}(T)$. \textbf{d,} $M(T)$ (blue) and $\frac{dM(T)}{dT}$ (red) with a \SI{0.1}{T} magnetic field along the c-axis. \textbf{e,} $M(\mu_0H)$ at several temperatures with the same magnetic field direction. \textbf{f,} $M(T)$ at \SI{1}{T} from the $M(\mu_0H)$ measurements.}
    \label{fig:CrTe800}
\end{figure*}

Since PXRD measurements revealed small amounts of CrTe in the higher $x$ \mct\, we grew phase-pure CrTe crystals using the same Sb-flux method (spinning at 650$^\circ$C) as the \mct\ crystals. Furthermore, due to the complex Cr-Te binary phase diagram~\cite{Cr_Te_binaryphasediagram} and the variety of reported Cr$_x$Te$_y$ phases with $x \approx y$ (see, for example, Refs.~\cite{cr3te4, cr2p76cr4, cr0p87te}) we also grew CrTe single crystals using the same recipe but only cooling to 800$^\circ$C at which point the growth was spun to further explore this phase space. These batches of CrTe had similar PXRD patterns as shown in Fig.~\ref{fig:CrTe_PXRD} where only CrTe and Sb could be detected. Furthermore, both batches showed similar magnetic and magnetotransport behaviors as shown in Fig.~\ref{fig:CrTe650} for the crystals spun at 650$^\circ$C and in Fig.~\ref{fig:CrTe800} for the crystals spun at 800$^\circ$C. In both cases below-room-temperature soft magnetic ordering is observed (276 and \SI{289}{K}, respectively), as well as a non-monotonic $\sigma_{xy}^{AH}(T)$ (determined by linearly fitting the high field $\rho_{xy}(\mu_0H)$ and extrapolating to zero magnetic field) and a monotonic saturated magnetization (taken at \SI{1}{T}). These results decisively show that this CrTe phase can not be responsible for the results shown in the main paper.

\subsection*{Supplementary Note 3: Energy-dispersive X-ray Spectroscopy}

\begin{figure*}
    \centering
    \includegraphics[width=0.8\textwidth]{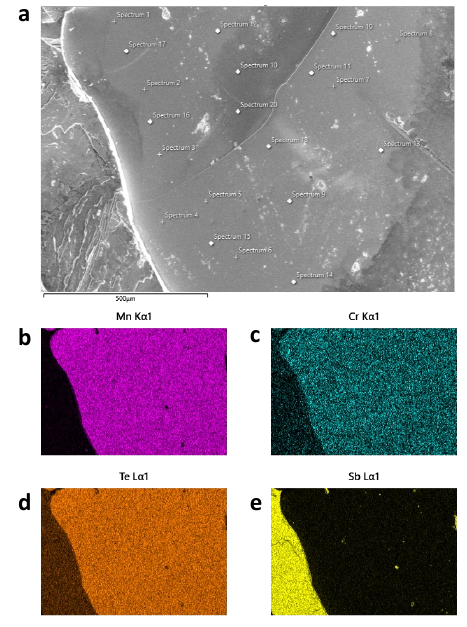}
    \caption{\textbf{EDX on $x=0.10$.} \textbf{a,} SEM image of a crystal of \mct\ with $x=0.10$, the highest doping accessed in this study. To the left side of the image is a chunk of residual flux. The scale bar on the bottom left is \SI{0.5}{mm}. White crosses represent regions measured numerically; these results are presented in ~\ref{tab:edx}. \textbf{b, c, d, e,} EDX maps of Mn, Cr, Te, and Sb respectively. While the Mn, Cr, and Te signals show up relatively uniformly in the crystal, the Sb signal is confined to the residual flux.}
    \label{fig:edx_p1}
\end{figure*}

\begin{table}[h]
\centering
\begin{tabular}{|l|l|l|l|}
\hline
Label & Mn & Cr & Te \\
\hline
Spectrum 1  & 47.41 & 5.12 & 47.47 \\
Spectrum 2  & 47.09 & 5.11 & 47.80 \\
Spectrum 3  & 46.74 & 5.23 & 48.03 \\
Spectrum 4  & 47.16 & 5.28 & 47.56 \\
Spectrum 5  & 46.73 & 5.40 & 47.87 \\
Spectrum 6  & 46.86 & 5.10 & 48.03 \\
Spectrum 7  & 46.94 & 4.88 & 48.19 \\
Spectrum 8  & 47.21 & 4.96 & 47.83 \\
Spectrum 9  & 46.44 & 5.46 & 48.10 \\
Spectrum 10 & 46.85 & 5.05 & 48.10 \\
Spectrum 11 & 47.30 & 4.95 & 47.75 \\
Spectrum 12 & 47.27 & 4.86 & 47.86 \\
Spectrum 13 & 47.17 & 4.87 & 47.95 \\
Spectrum 14 & 47.34 & 4.81 & 47.85 \\
Spectrum 15 & 47.55 & 4.85 & 47.59 \\
Spectrum 16 & 47.60 & 4.74 & 47.65 \\
Spectrum 17 & 47.29 & 5.00 & 47.71 \\
Spectrum 18 & 46.99 & 5.18 & 47.83 \\
Spectrum 19 & 47.00 & 4.75 & 48.25 \\
Spectrum 20 & 47.07 & 4.96 & 47.97 \\
\hline
Average & 47.10 & 5.03 & 47.87 \\
Standard Deviation & 0.29 & 0.21 & 0.21 \\
\hline
\end{tabular}
\caption{Elemental composition in atomic \% measured from EDX spectra for the $x=0.10$ sample shown in \ref{fig:edx_p1}.}
\label{tab:edx}
\end{table}

Fig.~\ref{fig:edx_p1}a shows a scanning electron microscope (SEM) image of a crystal of \mct\ with $x=0.10$, the highest doping accessed in this study. To the left side of the image is a chunk of residual flux. The scale bar on the bottom left is \SI{0.5}{mm}. Panels b-e present EDX maps of Mn, Cr, Te, and Sb respectively. While the Mn, Cr, and Te signals show up relatively uniformly in the crystal, the Sb signal is confined to the residual flux. The white crosses in panel a represent regions measured numerically; these results are presented in Table~\ref{tab:edx}. Over the 20 spots measured on this crystal the doping is quite homogeneous: the standard deviation on the Cr concentration is only about 4\% of its average value. This small standard deviation is consistent throughout the \mct\ series in the measured concentration range.

\subsection*{Supplementary Note 4: $\frac{d\rho_{xx}}{dT}$ and RMCD}

\begin{figure*}
    \centering
    \includegraphics[width=0.9\textwidth]{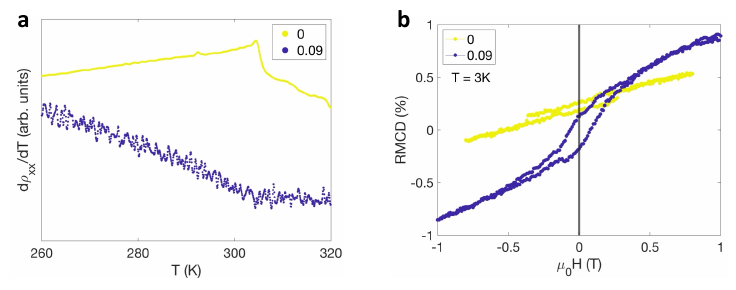}
    \caption{\textbf{$\frac{d\rho_{xx}}{dT}$ and RMCD.} \textbf{a,} $\frac{d\rho_{xx}}{dT}$ as a function of temperature near $T_N$ for samples of MnTe and $x = 0.09$. \textbf{b,} RMCD measurements as a function of magnetic field at \SI{3}{K} for samples of MnTe and $x = 0.09$.}
    \label{fig:drhodTandRMCD}
\end{figure*}

Fig.~\ref{fig:drhodTandRMCD}a shows $\frac{d\rho_{xx}}{dT}$ as a function of temperature near $T_N$ for $x = 0$ and 0.09 samples. Clear anomalies can be observed at $T_N$. Panel b presents RMCD measurements on samples with the same $x$ at \SI{3}{K}. As mentioned in the main text, $x =0.09$ has a much more pronounced hysteresis loop without steps, indicating the sample is reasonably homogeneous on the 1-2 micron scale of the measurement.

\subsection*{Supplementary Note 5: Out-of-plane magnetic moment reproducibility}

\begin{figure*}
    \centering
    \includegraphics[width=0.9\textwidth]{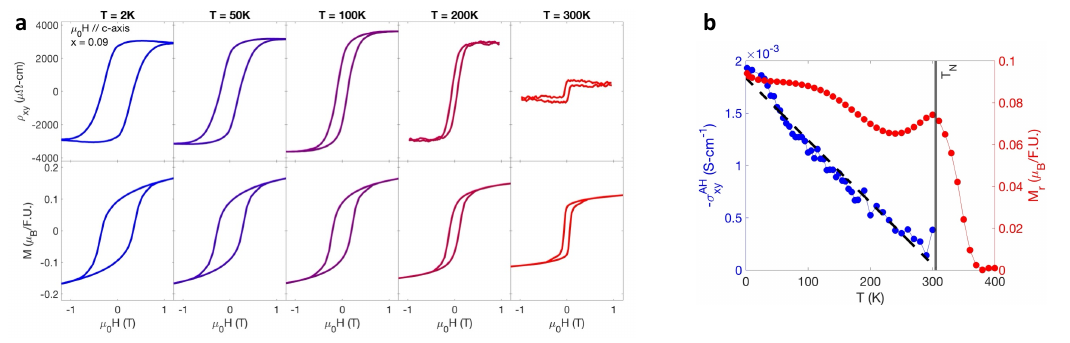}
    \caption{\textbf{Domain control with an out-of-plane magnetization --- second sample.} \textbf{a,} $\rho_{xy}$ (top) and $M$ (bottom) as a function of out-of-plane magnetic field for a sample of $x=0.09$ at several temperatures. \textbf{b,} $-\sigma_{xy}^{AH}$ and $M(\mu_0H=0^+)$ as a function of temperature extracted from the measurements shown in panel a. A linear fit to $\sigma_{xy}^{AH}$ is shown as in the main text.}
    \label{fig:OOPdomaincontrol_S1}
\end{figure*}

Fig.~\ref{fig:OOPdomaincontrol_S1} presents out-of-plane magnetic field AHE and magnetization data on another $x = 0.09$ sample, highlighting the reproducibility of the in-plane domain control presented in the main paper. Panel a shows $\rho_{xy}$ (top) and $M$ (bottom) as a function of out-of-plane magnetic field for a sample of $x=0.09$ at several temperatures. Like the sample shown in the main paper, the magnetization loop remains large near \SI{300}{K} whereas the AHE loop is substantially smaller than at \SI{2}{K}. Panel b shows $-\sigma_{xy}^{AH}$ and $M(\mu_0H=0^+)$ as a function of temperature extracted from the measurements shown in panel a. A linear fit to $\sigma_{xy}^{AH}$ is shown as in the main text. While the AHE measurements were not measured above \SI{300}{K}, a similar nearly linear dependence of $\sigma_{xy}^{AH}$ with decreasing temperature is measured below $T_N$ while the magnetization mostly grows at temperatures above $T_N$.

\subsection*{Supplementary Note 6: MnTe Relaxation}

\begin{figure*}
    \centering
    \includegraphics[width=0.9\textwidth]{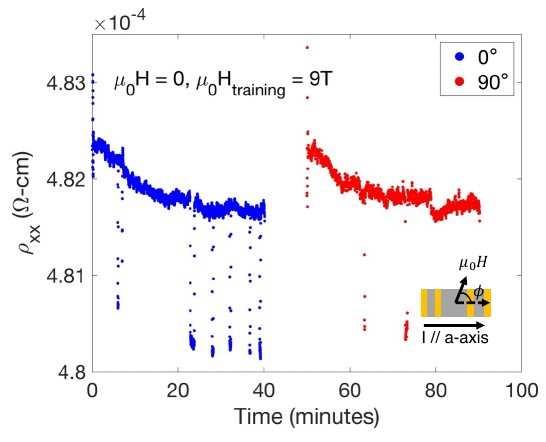}
    \caption{\textbf{MnTe domain relaxation.} $\rho_{xx}$ as a function of time at zero magnetic field after field-training the sample at \SI{9}{T} with magnetic field along the a-axis (blue) and perpendicular to this direction (red).}
    \label{fig:MnTerelaxation}
\end{figure*}

Fig.~\ref{fig:MnTerelaxation} presents $\rho_{xx}$ of MnTe as a function of time at zero magnetic field after field-training the sample at \SI{9}{T} with magnetic field along the a-axis (blue) and perpendicular to this direction (red). The exponential decay of the resistance as a function of time is evidence of the partial field detwinning of the altermangetic domains in the sample at high magnetic fields slowly being lost. This is in stark contrast to the \mct\ with a ferromagnetic moment in which the domains are pinned by the permanent moments.

\subsection*{Supplementary Note 7: In-plane magnetic moment reproducibility}

\begin{figure*}
    \centering
    \includegraphics[width=0.9\textwidth]{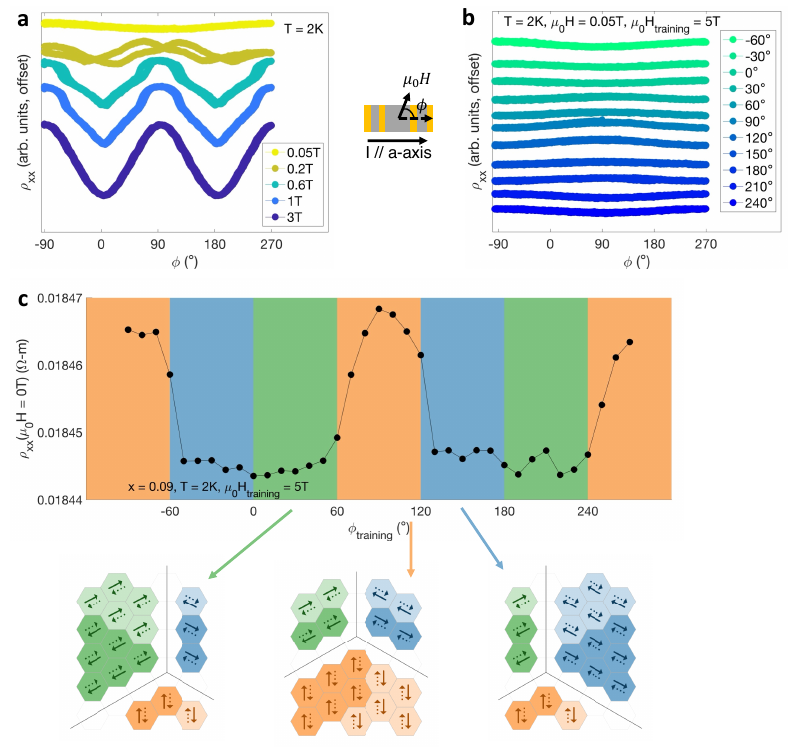}
    \caption{\textbf{Domain control with an in-plane magnetization --- second sample with current parallel to the a-axis.} \textbf{a,} The AMR of an $x = 0.09$ sample at \SI{2}{K} as an in-plane field of fixed magnitude is rotated. \textbf{b,} The AMR of this sample at \SI{0.05}{T} after training the sample with a \SI{5}{T} magnetic field at different angles $\phi_{training}$. \textbf{c,} Zero-field $\rho_{xx}$ of an $x=0.09$ sample as a function of $\phi_{training}$ ($\phi$ is the in-plane angle defined schematically in this figure) after applying and removing a \SI{5}{T} magnetic field at $\phi_{training}$ at \SI{2}{K}. Note the features every 60$^\circ$ which are highlighted by color shading.}
    \label{fig:IPdomaincontrol_S2}
\end{figure*}

Fig.~\ref{fig:IPdomaincontrol_S2} presents in-plane magnetic field AMR data at \SI{2}{K} on another $x = 0.09$ sample with the current applied parallel to the a-axis, highlighting the reproducibility of the in-plane domain control presented in the main paper. Panel a shows the AMR of this sample at different fixed magnetic fields and the same evolution from a onefold to a twofold pattern is observed with increasing field. Panel b shows the low field (\SI{0.05}{T}) AMR of this sample upon being trained with a \SI{5}{T} magnetic field at different angles $\phi_{training}$; once again, the magnetic field remanent magnetization always settles perpendicular to the current. Panel c shows the zero-field resistivity as a function of $\phi_{training}$ after applying a \SI{5}{T} training field; a qualitatively consistent pattern as the sample detailed in the main paper is observed.

\begin{figure*}
    \centering
    \includegraphics[width=0.9\textwidth]{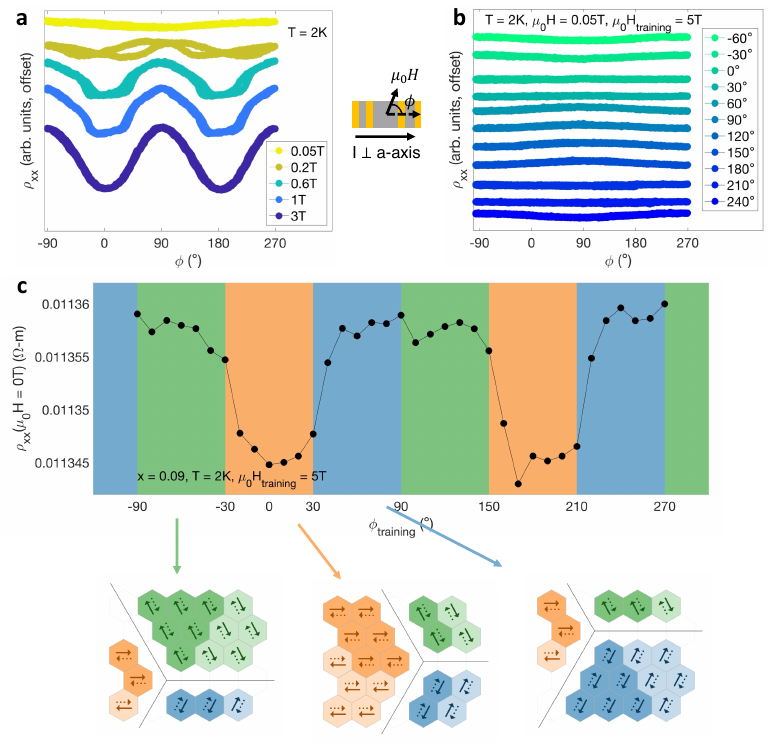}
    \caption{\textbf{Domain control with an in-plane magnetization --- sample with current perpendicular to the a-axis.} \textbf{a,} The AMR of an $x = 0.09$ sample at \SI{2}{K} as an in-plane field of fixed magnitude is rotated. \textbf{b,} The AMR of this sample at \SI{0.05}{T} after training the sample with a \SI{5}{T} magnetic field at different angles $\phi_{training}$. \textbf{c,} Zero-field $\rho_{xx}$ of an $x=0.09$ sample as a function of $\phi_{training}$ ($\phi$ is the in-plane angle defined schematically in this figure) after applying and removing a \SI{5}{T} magnetic field at $\phi_{training}$ at \SI{2}{K}. Note the features every 60$^\circ$ which are highlighted by color shading; they are 30$^\circ$ offset from those observed in the samples with current parallel to the a-axis.}
    \label{fig:IPdomaincontrol_perpI}
\end{figure*}

Fig.~\ref{fig:IPdomaincontrol_perpI} presents in-plane magnetic field AMR data at \SI{2}{K} on another $x = 0.09$ sample, this time with the current applied perpendicular to the a-axis. Panels a and b are qualitatively similar to the data presented in panels a and b of Fig.~\ref{fig:IPdomaincontrol_S2}. Once again, the low-field AMR shows that the remnant magnetization aligns with the short axis of the sample, indicating that this settling is not related to the crystallographic orientation of the sample. Panel c shows the zero-field resistivity as a function of $\phi_{training}$ after applying a \SI{5}{T} training field. The measured pattern is shifted relative to the samples with current applied along the a-axis as expected (note that the domains can be thought about by having been shifted by 30$^\circ$. The results are consistent with with the samples measured in the other orientation: when the altermagnetic domain with the moments aligned with the current are maximized, this leads to a lower resistance state.



\end{document}